\documentclass[prd,twocolumn,showpacs,floatfix,amsmath,nofootinbib,amssymb,floatfix]{revtex4}
\usepackage{graphicx,color,dcolumn,booktabs,bm}
\usepackage{longtable,lscape}
\usepackage{txfonts}
\usepackage{overpic}
\usepackage{amssymb}
\usepackage{indentfirst}
\usepackage{feynmf}   
\usepackage{slashed}  
\usepackage{cases}
\usepackage{color}
\usepackage{multirow}
\usepackage{epstopdf}
\usepackage{graphicx,color,dcolumn,booktabs,bm}
\usepackage[colorlinks,
            citecolor=blue,
            anchorcolor=red,
            menucolor=red,
            linkcolor=red,
            filecolor=red,
            runcolor=red,
            urlcolor=blue,
            frenchlinks=red]{hyperref}

\begin{document}

\title{Exploring open-charm decay mode $\Lambda_c\bar{\Lambda}_c$ of charmonium-like state $Y(4630)$}
\author{Xuewen Liu$^{1}$}
\email{liuxuewen@mail.nankai.edu.cn}
\author{Hong-Wei Ke$^2$}
\email{khw020056@tju.edu.cn}

\author{Xiang Liu$^{3,4}$}
\email{xiangliu@lzu.edu.cn}

\author{Xue-Qian Li$^{1}$}
\email{lixq@nankai.edu.cn}

\affiliation{$^1$School of Physics, Nankai University, Tianjin 300071, China \\
             $^2$School of Science, Tianjin University, Tianjin 300072, China\\
             $^3$School of Physical Science and Technology, Lanzhou University, Lanzhou 730000, China\\
             $^4$Research Center for Hadron and CSR Physics, Lanzhou University \& Institute of Modern Physics of CAS, Lanzhou 730000, China}

\begin{abstract}
The newly observed $X, Y, Z$ exotic states are definitely not in the standard 
$Q\bar Q'$ structures, thus their existence
composes a challenge to our understanding on the fundamental principles of hadron physics. Therefore the 
studies on their decay patterns which are determined by the non-perturbative QCD will definitely shed light on the concerned  physics. Generally the four-quark states might
be in a molecular state or tetraquark or their mixture. In this work, we adopt the suggestion that $Y(4630)$ is a charmonium-like tetraquark made of
a diquark and an anti-diquark. If it is true, its favorable decay mode should be $Y(4630)$ decaying into
an open-charm baryon pair, since such a transition occurs via strong interaction and is super-OZI-allowed.
In this work, we calculate the decay width of
$Y(4630)\to\Lambda_c\bar{\Lambda}_c$ in the framework of the quark
pair creation (QPC) model. Our numerical
results on the partial width computed in the tetraquark configuration coincide  with the Belle data within a certain error tolerance.

\end{abstract}
\maketitle

\section{Introduction}
\label{intro}

In 2007, the Belle collaboration reported that a $J^{PC}=1^{--}$
resonance peak $Y(4630)$ with mass $M=4634^{+9}_{-11}$ MeV and
width $\Gamma=92^{+41}_{-32}$ MeV appeared at the invariant mass spectra of the $e^+e^-\to\Lambda^+_c\Lambda^-_c$ channel~\cite{Pakhlova:2008vn}.

Besides an interpretation that the observed $Y(4630)$ is the $5^3S_1$ charmonium state \cite{Badalian:2008dv,Segovia:2008ta},
there are many alternative suggestions for the observed peak, for example,
$Y(4630)$ was considered to be induced by a threshold effect  instead of
being a genuine resonance \cite{vanBeveren:2008rt}, then it was also interpreted as a molecular state
made of $\psi(2S)$ and $f_0(980)$ by another theoretical physics group \cite{Guo:2010tk}.
Among those proposals, the suggestion that $Y(4630)$ is a tetraquark state is more favorable \cite{Maiani:2014aja,Cotugno:2009ys}.
In Ref.~\cite{Maiani:2014aja}, the $Y(4630)$ is identified as the ground
state with its orbital angular momentum $L=1$. It is noted that the mass and width of $Y(4630)$ are consistent
within errors with those for the $Y(4660)$ state ($M=4652\pm 10\pm8$MeV,
$\Gamma=68\pm 11\pm1$MeV), which is found in the invariant mass spectrum
of $\psi(2S)\pi^+\pi^-$ by the Belle collaboration \cite{Wang:2007ea,Wang:2014hta}.  By analyzing the
$\Lambda_c\bar{\Lambda}_c$ and $\psi(2S)\pi^+\pi^-$ spectra, Cotugno
\textit{et al.}  suggested that the $Y(4630)$ and
$Y(4660)$ could be the same tetraquark state, and is the first radial excitation of the
$Y(4360)$ with $L=1$ \cite{Cotugno:2009ys}.

In fact, $Y(4630)$  as a $[cq]_{\bar 3}[\bar c \bar q]_3$ tetraquark would more likely decay into  charmed baryon-pair ~\cite{Guo:2010tk,Cotugno:2009ys},
and
the ratio $\mathcal{BR}(Y\to\Lambda_c\bar{\Lambda}_c)/\mathcal{BR}(Y\to\psi(2S)\pi^+\pi^-)=25\pm7$  \cite{Cotugno:2009ys} suggests that the double baryon decay mode $\Lambda_c\bar{\Lambda}_c$ is strongly preferred.

However, there definitely may exist other decay modes beside of the $\Lambda_c\bar{\Lambda}_c$ pair, such as $D\bar{D}$, $D\bar{D}^*$, $D^*\bar{D}^*$, $J/\psi\eta$, $\psi(2S)\eta$, etc.
Such processes occur via color rearrangements which in principle can be depicted by hadronic loops even though the propagators in the loops do not correspond to real color-singlet particles (see in text),  so they  suffer from a loop suppression.
Even though the most promising tetraquark candidate $Z(4430)^+$ decays  into the $[c\bar q][\bar{c}{q}]$ mode  $\psi(2S)\pi^+$ ~\cite{Mizuk:2009da, Chilikin:2013tch, Aubert:2008aa,Aaij:2014jqa} with a broad width $\Gamma=172 \pm 13$ MeV,
this case is very different from $Y(4630)$. Since its mass is below the $\Lambda_c\bar{\Lambda}_c$ threshold it would overwhelmingly decay into open-charmed mesons.
For $Y(4630)$ case, as its mass is above the double-baryon threshold, the strong decay of such tetraquark state is OZI-super allowed.
Therefore, following the suggestions given by other groups here we will assume that the decay mode $Y(4630)\to\Lambda_c\bar{\Lambda}_c$ would be dominant,
namely this partial width could be at the same order of the total width.

A tetraquark is assumed to be made of the diquark-antidiquark $[cq][\bar c\bar q]$, where $q$ is a light quark either $u$ or $d$ and
$[cq]$ resides in a color anti-triplet whereas $[\bar c\bar q]$ is in a color triplet (in later calculations we do not distinguish between $u$ and $d$ at all).
In this work, we suppose that $Y(4630)$ is a tetraquark in the dynamic picture
suggested by Brodsky \textit{et al.} \cite{Brodsky:2014xia}. In
the tetraquark  a diquark and an anti-diquark are bound together via the QCD confinement, but
are separated by a substantial distance once they are
created. Thus the $Y(4630)$ state can be considered as a two-body meson-like state.
The picture we adopt in this work is slightly different from that proposed by Maiani \textit{et al.} \cite{Maiani:2004vq,Maiani:2014aja}, where the authors studied the
tetraquark states by means of their spin structure of a Hamiltonian
formalism \cite{Lebed:2015sxa}, in fact, the two pictures are in principle consistent.
Under this assignment, we study the strong decay of  $Y(4630)$ by computing the width of  $Y(4630)\to\Lambda_c\bar{\Lambda}_c$ in the quark pair creation (QPC) model.
The corresponding reaction mechanism is that first the diquark-antidiquark bound state is dissociated into a ``free" diquark-antidiquark system and
a light quark-antiquark pair is created from the
vacuum, then the quark and anti-quark would join the diquark and antidiquark respectively to
constitute a baryon-anti-baryon pair. Indeed, this association can be viewed as that due to soft gluon emission a light-quark pair is created and the soft gluons tear off the diquark-antidiquark
bound state, then by absorbing light quark and antiquark respectively they transit into color singlet baryons.
Moreover,  $M_{\Lambda_c}+M_{\bar\Lambda_c}$ is only slightly
below 4630 MeV, so that a suppression induced by matching different momenta as appearing at similar hadronic
processes, does not exist. Surely the whole dissociation process is governed by non-perturbative QCD, so that
one needs to introduce a few phenomenological factors which can only be obtained by fitting available data.

The paper is organized as follows: after this introduction, we
calculate the rate of $Y(4630)$ decaying  into the
$\Lambda_c\bar{\Lambda}_c$ pair in section.~\ref{3p0}$\&$\ref{sec:lamwave} and perform a
numerical analysis in Sec.~\ref{numerical}.  The other decays of $Y(4630)$ are discussed in Sec. \ref{otherdecays}.
Sec. \ref{conclusion} is devoted to our summary.

\section{The $Y(4630)\to \Lambda_c\bar{\Lambda}_c$ strong decay}
\label{Decay}

In this work, we use the two-body wave function for the  diquark-antidiquark bound system
$Y(4630)$, since the constituents (diquark and antidiquark) are treated as
two point-like color sources. In this structure, the diquark $Qq$ of color-anti-triplet in the tetraquark is in analog to a heavy $\bar Q$
residing in a common meson $Q\bar Q$ while $\bar Q\bar q$ is similar to $Q$ by the same color configuration.

The spin wave functions of a $Y(J^{PC}=1^{--})$ state with
$L=1$ in the basis of $|S_{qc},S_{\bar q \bar c},S_{\rm total},L\rangle_{J=1}$ can be assigned in four distinct states as~\cite{Maiani:2014aja}
\begin{eqnarray*}
  Y_1 &=& |0,0,0,1\rangle_1, \\
  Y_2 &=& 1/\sqrt{2} (|1,0,1,1\rangle_1+|0,1,1,1\rangle_1),\\
  Y_3 &=& |1,1,0,1\rangle_1,\\
  Y_4 &=& |1,1,2,1\rangle_1.
\end{eqnarray*}

In the following, we present all the details of calculating $Y(4630)\to \Lambda_c\bar{\Lambda}_c$ in the QPC model.

\subsection{Implementation in the QPC model}
\label{3p0} The QPC model
\cite{Micu,yaouanc,yaouanc-1,yaouanc-book,Beveren,BSG,sb} has been
widely applied to calculate the rates of Okubo-Zweig-Iizuka (OZI) allowed strong
decays \cite{Blundell:1995ev,qpc-2,Burns:2014zfa,Capstick:1986bm,qpc-90,ackleh,Zou,liu,Close:2005se,lujie,xiangliu-2860,
xiangliu-heavy,Chen:2007xf,Li:2008mz}, which obviously compose the dominant contributions to
the total widths of the concerned
hadrons.

\begin{center}
\begin{figure}[htb]
\scalebox{0.48}{\includegraphics{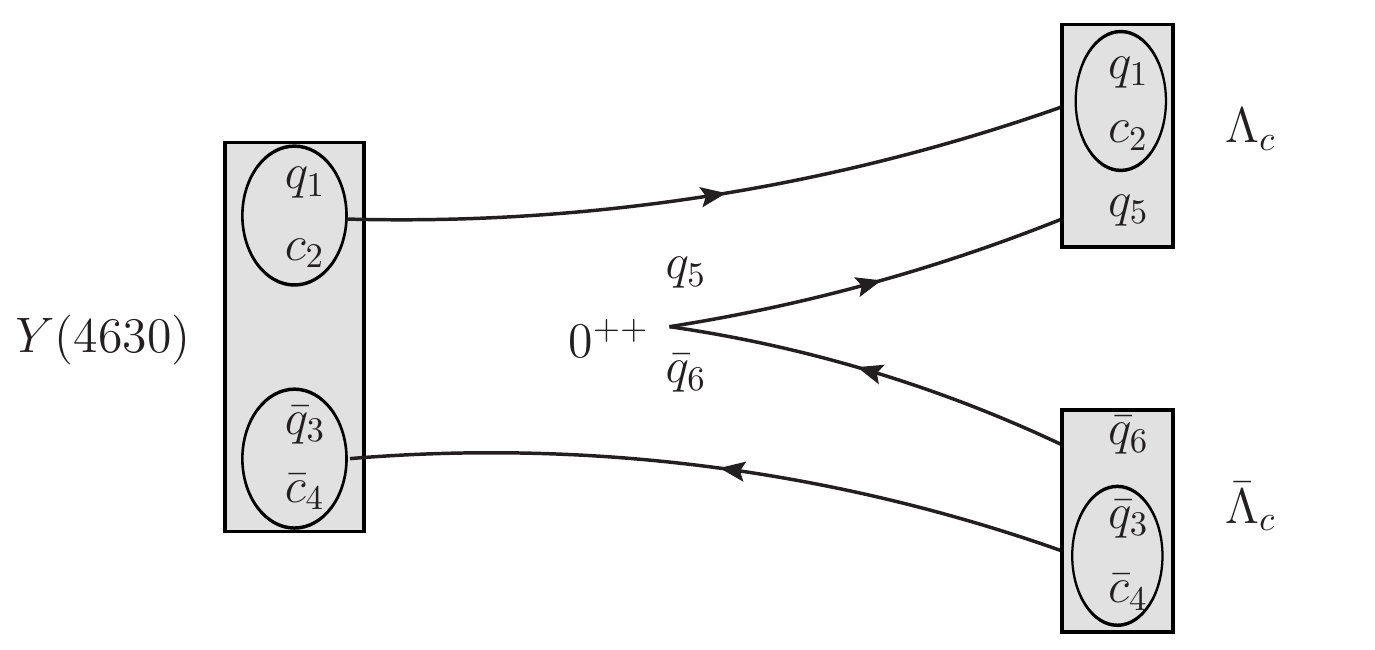}} \caption{The QPC
mechanism for decay $Y(4630)\to \Lambda_c+\bar{\Lambda}_c$, we
label the quark $c$ and antiquark $\bar c$ with subscripts 2 and 4,
as well understood. $q$ stands for the light quark $u/d$.}
\label{Fig:decay}
\end{figure}
\end{center}

As indicated in the introduction, we suppose $Y(4630)$ as a tetraquark in the diquark-antidiquark
structure, thus in our case, the decay of $Y(4630)$ is a dissociation process where the
diquark and antidiquark bound state is loosened by a quark-antiquark pair which
is created in vacuum. Concretely, the quark and antiquark of the pair
excited out from the vacuum would join the diquark and antidiquark respectively to compose a
$\Lambda_c\bar{\Lambda}_c$ pair, and the process is graphically shown
in Fig.~\ref{Fig:decay}.

The quantum number of the created quark pair is
$0^{++}$ \cite{Micu,yaouanc}. In the non-relativistic limit, the
transition operator is expressed as
\begin{eqnarray}
T&=& - 3 \gamma \sum_m\: \langle 1\;m;1\;-m|0\;0 \rangle\,
\int\!{\rm d}{\textbf{k}}_5\; {\rm d}{\textbf{k}}_6
\delta^3({\textbf{k}}_5+{\textbf{k}}_6) \nonumber\\&&
\times{\cal Y}_{1m}\left(\frac{{\textbf{k}}_5-{\textbf{k}_6}}{2}\right)\;
\chi^{56}_{1, -\!m}\; \varphi^{5 6}_0\;\,
\omega^{5 6}_0\; d^\dagger_{5i}({\textbf{k}}_5)\;
b^\dagger_{6j}({\textbf{k}}_6)\,, \label{tmatrix}
\end{eqnarray}
where $i$ and $j$ are the SU(3)-color indices of the created quark
and anti-quark. $\varphi^{56}_{0}=(u\bar u +d\bar d +s \bar
s)/\sqrt 3$ and $\omega_{0}^{56}=\delta_{ij}$ are for flavor and
color singlets, respectively. $\chi_{{1,-m}}^{56}$ is a spin triplet. Here the indices 5 and 6 distinguish
between the quark and antiquark respectively as shown in Fig.~\ref{Fig:decay}.
$\mathcal{Y}_{\ell m}(\mathbf{k})\equiv
|\mathbf{k}|^{\ell}Y_{\ell m}(\theta_{k},\phi_{k})$ denotes the
$\ell$th solid harmonic polynomial. $\gamma$ is a dimensionless
constant for the strength of quark pair creation from
vacuum and is fixed by fitting data.

In the dynamical picture of tetraquark, the (anti)diquark
is considered to be a point-like color source, then the two-body wave function(meson-like)
should be a good approximation to describe the inner structure of $Y(4630)$. Including the color ($\omega_Y^{[12][34]}$),
spin ($\chi_Y^{[12][34]}$), flavor ($\varphi_Y^{[12][34]}$) and the spatial
($\Psi_{n_{Y}L_{Y} M_{L_{Y}}}\left(\mathbf{p}_1,\mathbf{p}_2\right)$) parts, the wave function is written as

\begin{eqnarray}\label{eq:y}
&&\left|Y(n_{Y} \mbox{}^{2S_{Y}+1}L_{Y} \,\mbox{}_{J_{Y} M_{J_{Y}}})
({\textbf{K}}_{Y}) \right\rangle \nonumber\\&&= \sqrt{2
E_Y}\sum_{M_{L_{Y}},M_{S_{Y}}} \left\langle L_{Y} M_{L_{Y}} S_{Y} M_{S_{Y}} |
J_{Y} M_{J_{Y}} \right\rangle \nonumber\\&&
\quad\times\int \rm d
\mathbf{p}_1\rm d
\mathbf{p}_2\rm
\delta^3\left(\textbf{K}_{Y}-\mathbf{p}_1-{\mathbf{p}}_2\right) 
\Psi_{n_{Y}L_{Y} M_{L_{Y}}}\left(\mathbf{p}_1,\mathbf{p}_2\right)\nonumber\\
&&\quad\times
\chi_{S_{Y} M_{S_{Y}}}\varphi^{[1 2] [3 4]}_{Y}\omega^{[1 2] [3 4]}_{Y}
\left|\;[q_1\;q_2]\left(\mathbf{p}_1\right)
[\bar{q}_3\bar{q}_4]\left(\mathbf{p}_2\right)\right\rangle,
\end{eqnarray}
where we use the (super)subscript 1$\sim$4 to mark the (anti)quark
in the tetraquark as clearly shown in Fig~\ref{Fig:decay}. ${\textbf{K}}_Y$ is the 3-momentum of
$Y(4630)$, ${\textbf{p}}_{1(2)}$ is the 3-momentum of the
(anti)diquark.
$\mathbf{S}_Y=\mathbf{S}_{[q_1 q_2]}+\mathbf{S}_{[\bar
q_3 \bar q_4]}$ is the total spin.
$\mathbf{J}_Y=\mathbf{L}_Y+\mathbf{S}_Y$ denotes the total angular
momentum of $Y(4630)$.

We also consider the diquark-quark picture \cite{Anselmino:1992vg,Santopinto:2004hw,Ferretti:2011zz,Santopinto:2014opa,Gutierrez:2014qpa} for the $\Lambda_c$ baryon in where the internal degrees of freedom of the diquark are neglected as in the tetraquarks, then we have
\begin{eqnarray}\label{eq:baryon}
&&\left|\Lambda_c(M_{S_{\Lambda_c}})({\textbf{K}}_{\Lambda_c}) \right\rangle=\sqrt{2
E_{\Lambda_c}} \int \rm d
\mathbf{p}_1\rm
d\mathbf{k}_3\rm
\delta^3\left(\textbf{K}_{\Lambda_c}-\mathbf{p}_1-{\mathbf{k}}_3\right)\nonumber\\
&&\times \Psi_{\Lambda_c}\left(\mathbf{p}_1,\mathbf{k}_3\right)
\chi_{\frac{1}{2},M_{S_{\Lambda_c}}} \varphi^{[1 2]3}_{\Lambda_c}\omega^{[1 2]3}_{\Lambda_c}\nonumber\\
&&\times\left|\;[q_1 q_2]\left(\mathbf{p}_1\right)q_3\left(\mathbf{k}_3\right)\right\rangle,
\end{eqnarray} 
where the (super)subscripts in the expressions correspond to the  constituent quark and the diquark, and ${\textbf{K}}_{\Lambda_c}$ is the 3-momentum of ${\Lambda_c}$, ${\textbf{p}}_{1}({\textbf{k}}_{3})$ is the 3-momentum of the
diquark(quark).
The quantum numbers of $\Lambda_c$ are known as $J^P=\frac{1}{2}^+$ and $L=0$, so we only use  $M_{S_{\Lambda_c}}$($=M_{J_{\Lambda_c}}$) to label the spin projection state.

The wave functions respect the normalization conditions
\begin{eqnarray}\label{eq:nor-cond}
\langle Y(\textbf{K}_Y)|Y(\textbf{K}'_Y) \rangle &=& 2E_Y\,
\,\delta^3(\textbf{K}_Y-\textbf{K}'_Y),\;\\
\langle {\Lambda_c}(\textbf{K}_{\Lambda_c})|{\Lambda_c}(\textbf{K}'_{\Lambda_c})
\rangle &=& 2E_{\Lambda_c}\,
\,\delta^3(\textbf{K}_{\Lambda_c}-\textbf{K}'_{\Lambda_c}).\;\\\nonumber
\end{eqnarray}

For $Y(4630)\to \Lambda_c+\bar{\Lambda}_c$ process, the transition hadronic matrix element is written as
\begin{eqnarray*}
\langle
\Lambda_c\bar{\Lambda}_c|S|Y(4630)\rangle&=&I-i2\pi\delta(E_f-E_i)
\langle\Lambda_c\bar{\Lambda}_c|T|Y(4630)\rangle.
\end{eqnarray*}
In the center of the mass frame of $Y(4630)$, $\textbf{K}_Y=0$
and $\textbf{K}_
{\Lambda_c}=-\textbf{K}_{\bar{\Lambda}_c}=\textbf{K}$. Then, we have
{\small \begin{eqnarray}\label{T-matrix}
&&\langle\Lambda_c\bar{\Lambda}_c|T|Y(4630)\rangle
= -3\gamma\sqrt{8 E_Y E_{\Lambda_c}
E_{\bar{\Lambda}_c}}\;\;\nonumber\\&&\quad\times
\sum_{\renewcommand{\arraystretch}{.1}\begin{array}[t]{l}
\scriptstyle M_{L_Y},M_{S_Y},m
\end{array}}\sum_{\renewcommand{\arraystretch}{.1}\begin{array}[t]{l}
\scriptstyle M_{S_{\Lambda_c}},M_{S_{\bar{\Lambda}_c}}
\end{array}}\renewcommand{\arraystretch}{1}
\langle 1\;m;1\;-m|\;0\;0 \rangle \nonumber\\&&\quad\times
\langle s_5\;m_5;s_6\;m_6|\;1\;-m \rangle
\langle
L_Y M_{L_Y} S_Y M_{S_Y} | J_Y M_{J_Y} \rangle\nonumber\\&&\quad\times\langle
S_{12} M_{S_{12}}S_{34} M_{S_{34}} | S_Y M_{S_Y} \rangle
\langle S_{12} M_{S_{12}}S_5 M_{S_5} | S_{\Lambda_c} M_{S_{\Lambda_c}} \rangle\nonumber\\&&\quad\times
\langle S_{\Lambda_c} M_{S_{\Lambda_c}} 00| J_{\Lambda_c} M_{J_{\Lambda_c}} \rangle
\langle S_{34} M_{S_{34}}S_6 M_{S_6} | S_{\bar{\Lambda}_c} M_{S_{\bar{\Lambda}_c}} \rangle\nonumber\\&&\quad\times
\langle S_{\bar{\Lambda}_c} M_{S_{\bar{\Lambda}_c}} 00| J_{\bar{\Lambda}_c} M_{J_{\bar{\Lambda}_c}} \rangle
\langle\varphi^{[12]5}_{\Lambda_c} \varphi^{[34]6}_{\bar{\Lambda}_c} | \varphi^{[12][34]}_Y \varphi^{56}_0
 \rangle\nonumber\\&&\quad\times
 \langle \omega^{[12]5}_{\Lambda_c}\omega^{[34]6}_{\bar{\Lambda}_c} | \omega^{[12][34]}_Y \omega^{56}_0 \rangle
I^{M_{L_Y},m}({\textbf{K}}) \;.
\end{eqnarray}}

The expressions of Eq.~(\ref{T-matrix}) for $Y_{1,2,3,4}$ states are explicitly written out in terms of $I^{M_{L_Y},m}(\textbf{K})$ as listed in the Appendix~\ref{appb}. The spatial integral $I^{M_{L_Y},m}(\textbf{K})$ manifests an overlap between the spacial parts of the initial state (including the created light quark
pair) and the final state, and is expressed as
\begin{eqnarray}
&&I^{M_{L_Y},m}(\textbf{K}) = \int\!\rm d\mathbf{p}_1\rm d\mathbf{p}_2\rm d\mathbf{k}_5\rm d\mathbf{k}_6\,\nonumber\\
&&\quad\times\delta^3(\mathbf{p}_1+\mathbf{p}_2)\delta^3(\textbf{K}_{\Lambda_c}-\mathbf{p}_1-\mathbf{k}_5)
\delta^3(\textbf{K}_{\bar{\Lambda}_c}-\mathbf{p}_2-\mathbf{k}_6)
\nonumber\\&&\quad\times\delta^3(\mathbf{k}_5+\mathbf{k}_6)
\Psi^*_{{\Lambda_c}}(\mathbf{p}_1,\mathbf{k}_5)\Psi^*_{{\bar{\Lambda}_c}}(\mathbf{p}_2,\mathbf{k}_6)\nonumber\\&&\quad\times
\Psi_{n_Y L_Y M_{L_Y}}(\frac{\mathbf{p}_1-\mathbf{p}_2}{2})
\mathcal{Y}_{1m}\Big(\frac{\mathbf{k}_5-\mathbf{k}_6}{2}\Big)\nonumber\\&&\quad =
\int\!\rm d\mathbf{p}
\Psi^*_{{\Lambda_c}}(\mathbf{p}-\mu\mathbf{K})\Psi^*_{{\bar{\Lambda}_c}}(-\mathbf{p}+\nu\mathbf{K})\nonumber\\&&\quad\times
\Psi_{n_Y L_Y M_{L_Y}}(\mathbf{p})
\mathcal{Y}_{1m}\Big(\mathbf{p}-\mathbf{K}\Big),
\label{integral}
\end{eqnarray}
where $\mu=m_{[cq]}/(m_{[cq]}+m_q)$ and $\nu=m_{[\bar{c}\bar{q}]}/(m_{[\bar{c}\bar{q}]}+m_{\bar{q}})$.  Following the literature in this field, we
employ the simple harmonic oscillator (SHO) wavefunctions to stand for the spacial parts of the two-body wave functions of $Y(4630)$. Their explicit forms are collected in the
appendix~\ref{appa}. The wavefunction of $\Lambda_c$ will be considered in the next section.

With the transition amplitude given in Eq. (\ref{T-matrix}), the matrix element can be
rewritten in terms of the
helicity amplitude $\mathcal{M}^{M_{J_Y}M_{J_{\Lambda_c}}M_{J_{\bar{\Lambda}_c}}}$ as
\begin{eqnarray}
\langle\Lambda_c\bar{\Lambda}_c|T|Y(4630)\rangle=\delta^3(\mathbf{K}_{\Lambda_c}+
\mathbf{K}_{\bar{\Lambda}_c}-\mathbf{K}_Y)\mathcal{M}^{M_{J_{\Lambda_c}}M_{J_{\bar{\Lambda}_c}}}.
\end{eqnarray}
The decay width of  $Y(4630)\to \Lambda_c\bar{\Lambda}_c$ is then
\begin{eqnarray*}
\Gamma_Y=\pi^2\frac{|\mathbf{K}|}{M_Y^2}\frac{1}{2J_Y+1}
\sum_{\renewcommand{\arraystretch}{.5}\begin{array}[t]{l}
\scriptstyle M_{J_{M_{\Lambda_c}}},
\scriptstyle M_{J_{M_{\bar{\Lambda}_c}}}
\end{array}}
\Big|\mathcal{M}^{M_{J_{\Lambda_c}}M_{J_{\bar{\Lambda}_c}}}\Big|^2\, ,
\end{eqnarray*}
where $|\textbf{K}|$, as aforementioned, is the 3-momentum of
the final states in the center of mass frame.

\subsection{Baryon wavefunction}
\label{sec:lamwave}

The charmed baryon $\Lambda_c$ is considered as the  $[cq]$-$q$ picture in our scenario, then a two body wavefunction, which can be gained by solving the Schr\"odinger
equation, could be a reasonable approximation.

For our concrete calculation, we employ a non-relativistic  Cornell-like potential where the concerned free parameters are fixed by
fitting the mass spectra of charmed baryons. By solving the Schr\"odinger equation we obtain the wave function of $\Lambda_c$.
The general Hamiltonian of a diquark-quark system (i.e. a two body system)  can be written as
\begin{equation}\label{eq:Ham}
 H=\frac{{\bf p}^2_{[cq]}}{2m_{[cq]}}+m_{[cq]}+\frac{{\bf p}_q^2}{2m_q}+m_q+V(r) ,\\
\end{equation}
where the $m_{[cq]}({\bf p}_{[cq]})$ and $m_q({\bf p}_q)$ are the masses(3-momenta) of the diquark $[cq]$ and quark $q$ respectively.

It is worth of pointing out that in literature, the diquark-quark structure might be different, namely the two light quarks make a light diquark and the heavy quark stands as a color source.
Instead the baryon still might be in $[Qq]_{\bar 3}q_3$ structure \cite{Fleck:1988vm}, especially in our case the diquark (anti-diquark) does not have time to recombine into $Q_3[qq]_{\bar 3}$ by color rearrangement, namely the original diquark structure would remain to make a color singlet baryon by absorbing a light quark. The interaction potential is

\begin{equation}\label{eq:Vcon}
V(r)=-\frac{4}{3}\frac{\alpha_s}{r} + b r^\kappa+c,
\end{equation}
where  $-4/3$ is the color factor specific to  ${\mbox{\boldmath$3$}}$-${\mbox{\boldmath$\bar 3$}}$ attraction, $b$ is the string tension and $c$ is a global zero-point energy. Here we take the $b r^\kappa+c$ part as the confinement which is slightly different form the usual Cornell  $b r+c$ potential. $\alpha_s$ is the phenomenological strong coupling constant.

In this work, since only the wave function of $\Lambda_c^+(2286)$ which is in S-wave is needed,  the hyperfine interactions  including the spin-spin interaction, the spin-orbit interaction and the color tensor
interaction \cite{Capstick:1986bm} are not included. 

With the diquark mass $m_{[cq]}=$1.86 GeV  which is calculated by the QCD sum rules~\cite{Kleiv:2013dta} and the light constituent quark mass $m_{q}=$0.33 GeV, the parameters are fixed to be: $\alpha_s=0.45, b=0.135$ GeV$^2, \kappa=0.84, c=0.333$ GeV. Here, as theoretical inputs, we ignore possible inaccuracies of the parameters.

\begin{center}
\begin{figure}[htbp]
\scalebox{0.35}{\includegraphics{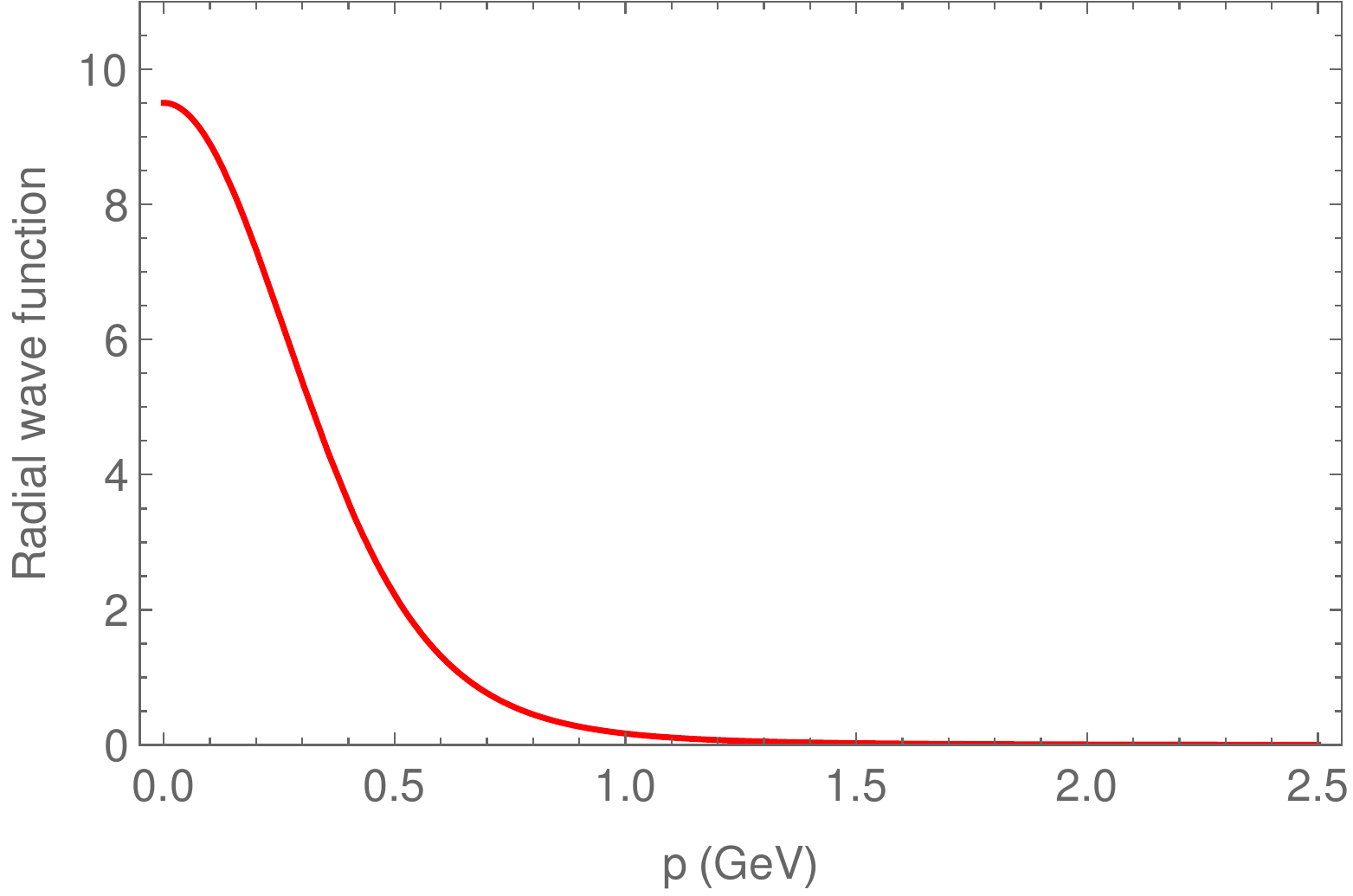}}
\caption{The radial wave function of  $\Lambda_c$ as a diquark-quark system.} \label{Fig:lamwave}
\end{figure}
\end{center}

The fitted spectra are presented in Table.~\ref{Tab:spectra},  and  a comparison with the experimental data and other theoretical predictions in literature are also listed in the table. The radial wave function of $\Lambda_c(2286)$ is plotted in Fig.~\ref{Fig:lamwave}.

\renewcommand{\arraystretch}{1.5}
\begin{table}[hbtp]
\caption{The fitted spectra of charmed baryons with different quantum numbers, including a comparison with the experimental data and other theoretical predictions in literature. Here, the masses of the baryons are in units of MeV.}\label{Tab:spectra}
\begin{tabular}{cccccc}
\toprule[1pt]\toprule[1pt]
States              & PDG \cite{Agashe:2014kda}   & This work  & Ref. \cite{Capstick:1986bm}   & Ref.\cite{Ebert:2011kk}   & Ref.\cite{Chen:2014nyo}  \\ \midrule[1pt]
$|1S,1/2^+\rangle$  & 2286.46 & 2286.1   & 2265    & 2286     & 2286             \\
$|2S,1/2^+\rangle$  & 2766.6  & 2768.5   & 2775    & 2769     & 2766       \\
$|3S,1/2^+\rangle$  &         & 3115.0   & 3170    & 3130     & 3112        \\
$|1P,1/2^-\rangle$  & 2592.3  & 2627.6   & 2630    & 2598     & 2591       \\
$|2P,1/2^-\rangle$  & 2939.3  & 3006.9   & 3030    & 2980     & 2989           \\
$|1D,5/2^+\rangle$  & 2881.53 & 2864.9   & 2910    & 2880     & 2879        \\
 \bottomrule[1pt]
\bottomrule[1pt]
\end{tabular}
\end{table}

\subsection{Numerical results}
\label{numerical}

Following Ref.~\cite{Godfrey:1986wj}, in the numerical computations, we adopt the model parameter $\gamma=6.3$  which is considered as universal in the QPC model. Meanwhile the $R$ value for P-wave tetraquark in the SHO wave function,  which represents the mean-square root (RMS) radius, can either not be determined from an underlying principle,
so we perform a numerical analysis dependent on $R_Y$ with certain ranges,
where $R_Y$ denote the $R$-value of the wave functions for $Y(4630)$ in tetraquark structure.
Since there still exists an ambiguity about the inner structure of $Y(4630)$,
we calculate the decay width for two possible cases: assuming (1) $Y(4630)$ as the
ground state with the radial quantum number $n_r=1$ and (2) the first radial
excitation with $n_r=2$ assignments.

\begin{center}
\begin{figure}[htbp]
\scalebox{0.24}{\includegraphics{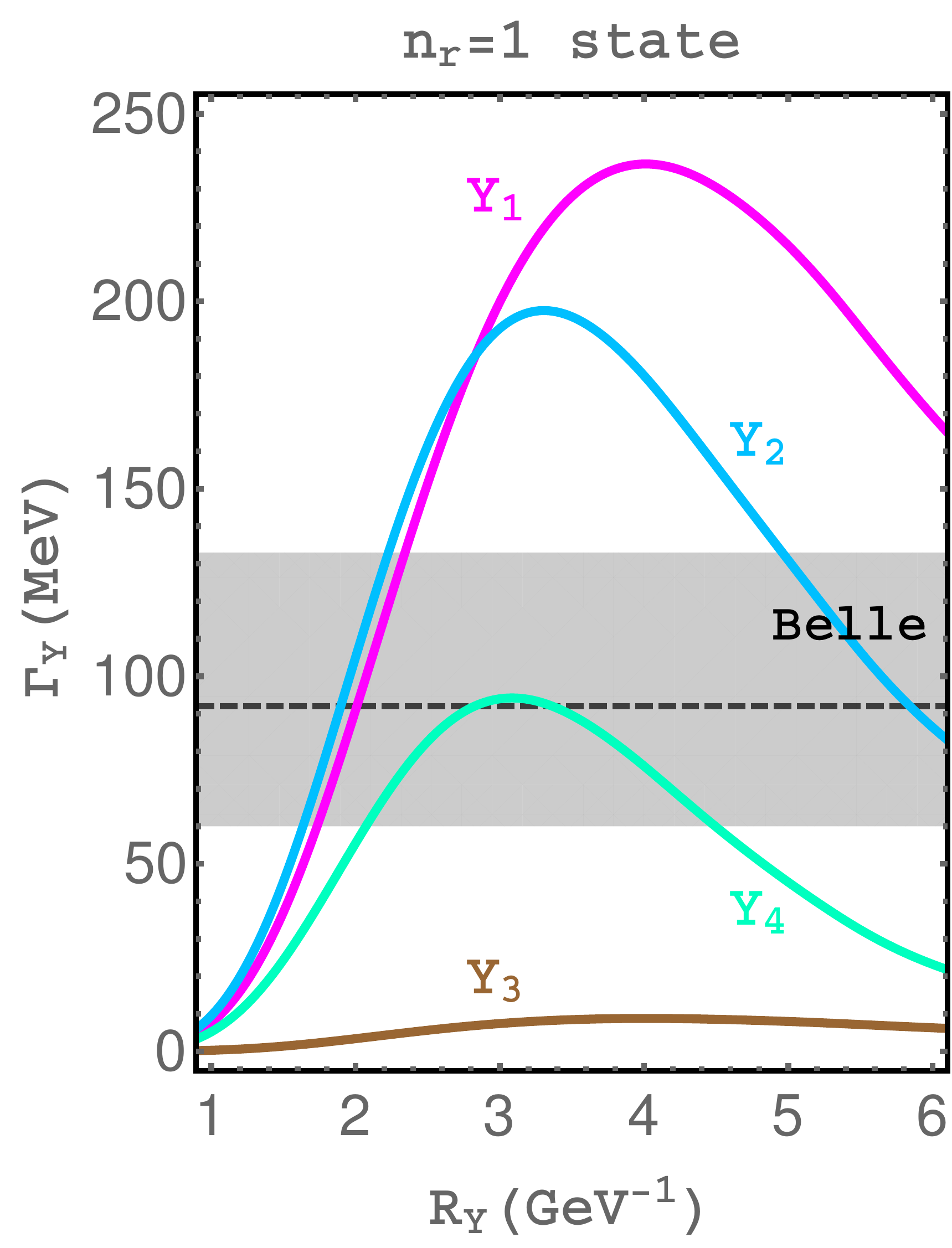}\includegraphics{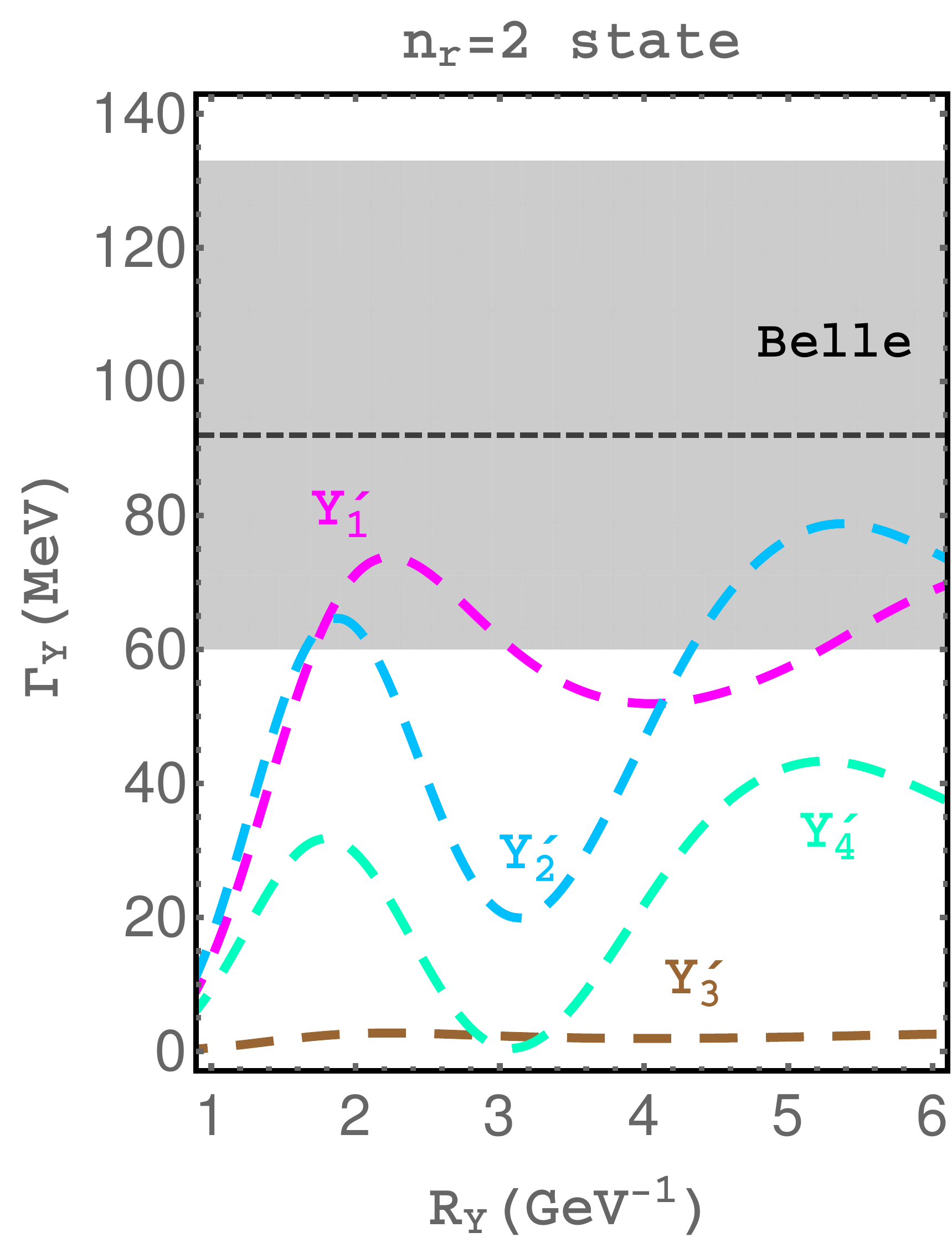}}
\caption{Dependence of the predicted partial width of $Y(4630)\to\Lambda_c\bar\Lambda_c$ on $R_Y$. The
Belle data are shown in the plot for a comparison. The black dashed line and the gray band correspond to the central value and
error for the total width of $Y(4630)$ measured by the Belle collaboration ($\Gamma=92^{+41}_{-32}$ MeV).
The colored curves correspond to the four different spin assignments
$Y_1, Y_2, Y_3, Y_4$ respectively. The solid and dashed curves correspond to the $n_r=1$ and $n_r=2$ cases.  Here, in the right panel we use the prime to distinguish $``Y"$ states in the two cases.} \label{Fig:results}
\end{figure}
\end{center}

We first compute the decay width of $Y(4630)\to \Lambda_c\bar{\Lambda}_c$ with the $n_r=1$ assignment.
The  left panel of Fig.~\ref{Fig:results} 
shows the  dependence of the calculated width $\Gamma_Y$ on $R_Y$   within a range
($1\sim6$)GeV$^{-1}$. The colored curves correspond to the four spin states $Y_{1,2,3,4}$ which are marked on the figures.
 As discussed before, the decay mode $Y(4630)\to\Lambda_c\bar{\Lambda}_c$ should be dominant, so we compare this calculated partial width with the total width of Y(4630).
In the plot one can find the predicted width for the $Y_{1,2,4}$ assignments do coincide with the data and the error band of 1$\sigma$ (gray region) given by the Belle collaboration ($\Gamma=92^{+41}_{-32}$ MeV).

For the $Y_3$ case the figure shows that the values of the curves are obviously lower than the data $\Gamma=92^{+41}_{-32}$ MeV.
This suppression is caused by the relatively small overlap between the spin wave functions of initial and final states (one can see the appendix~\ref{appb} for some details). Therefore it is concluded that
the data do not favor  $Y(4630)$ to be a ground state with $Y_3$ spin structure.

Next, as $Y(4630)$ being assigned as the first radial excitation state, 
our numerical results are shown at the right panel of Fig.~\ref{Fig:results} for  all the four spin assignments.
The results show that the $Y_{1,2}$ states can meet with the experimental data 
as long as $R_Y$ lies in a range of $1.5\sim3$ GeV$^{-1}$ and/or around 5GeV$^{-1}$.
The values correspond to the $Y_4$ state  are slightly lower, however,  they are still of the same order as the total width.
Again,  for $Y_3$ state, the situation is similar to that for $n_r=1$, the computed width are much below the data.

In a brief summary,  our numerical results indicate that within certain regions of the parameter $R_Y$,  the partial width of  $Y(4630)\to\Lambda_c\bar\Lambda_c$ can be comparable with the Belle data.  Given the fact that the peak of $Y(4630)$ has only been observed at the invariant spectrum of $\Lambda_c\bar\Lambda_c$, one is tempted to assume that the $\Lambda_c\bar\Lambda_c$ mode dominates the decay of $Y(4630)$ Moreover our calculation indicates that this predicted partial width is comparable with the total width of $Y(4630)$.
This consistency supports the assumption that the $Y(4630)$ is a P-wave  tetraquark in the diquark-antidiquark configuration and decays  mainly into double charmed baryons. We will make more discussions on this issue in the next section.

Also $Y(4630)$ could be in either the radial ground state with $n_r=1$ or the first excited state with $n_r=2$. In other words, the present data cannot rule out any of the two possible configurations. So definitely it needs to be studied with more experimental information in the future to decide the more accurate nature of $Y(4630)$, so as the spin structures.

\section{Discussions on other decay modes}
\label{otherdecays}

As discussed in the introduction, beside the  dominant $Y(4630)\to\Lambda_c\bar{\Lambda}_c$, there may exist other decay modes, such as $D\bar{D}$, $D\bar{D}^*$, $D^*\bar{D}^*$, $\psi(2S)\pi^+\pi^-$, $\psi(2S)\eta$, etc.
For instance, if one considers the both observed $Y(4630)$ and $Y(4660)$ to be tetraquark states~\cite{Cotugno:2009ys},  $Y(4630(4660))\to\psi(2S)\pi^+\pi^-$ occurs through a quark rearrangement process.

For the tetraquark structure, this decay mode requires a quark-antiquark rearrangement which is also a color exchange process. In the process
a  quark and an antiquark  which belong to different clusters are switched round
to produce the final states.

\begin{center}
\begin{figure}[htbp]
\scalebox{0.35}{\includegraphics{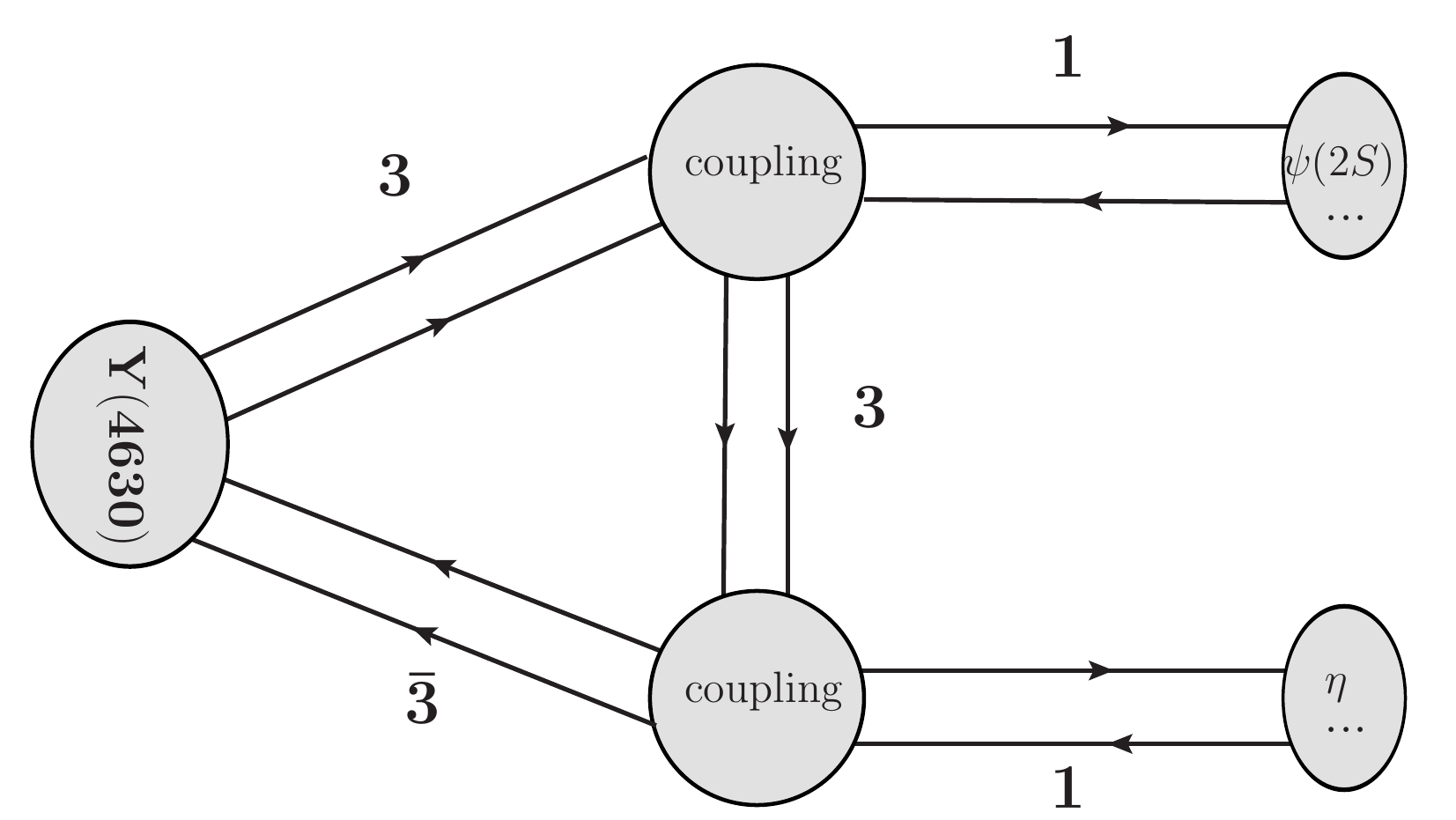}}
\caption{Effective diagram for the decay of $Y(4630)$ to $\psi(2S) \eta$, etc.} \label{Fig:decay3}
\end{figure}
\end{center}

In the figure \ref{Fig:decay3}, tracing the diquark (antiquark) flow lines, one can draw an effective hadronic Feynman diagram as a diquark (scalar or vector) which brings a color-content (color-triplet $3$ or anti-triplet ($\bar 3$)
is exchanged between the diquark $[cq]$ and antidiquark $[\bar c\bar q]$,  and results in the final state to be in color-singlet.

In figure, such processes occur via a hadronic loop, therefore is suffering from a loop suppression.
This Feynman diagram is similar to the final state interaction where all lines corresponding to (no matter inside the loop or outside finally produced hadrons) color-singlet hadrons, thus only difference between the quark rearrangement and final state interaction is their color configurations. But both of them are suppressed. In our another paper, we estimated the rates of $Y(4630)\to \Lambda_c\bar\Lambda_c\to p\bar p, D\bar D, D\bar D^*, \pi\pi, K^+K^-$ etc. through hadronic rescattering and found that such as final states could be observed by much more accurate measurements~\cite{Guo:2016iej}. Similarly, we may conjecture that the color- re-arrangement which proceeds along similar way should have comparable rates.

In fact, such quark exchange mechanism was investigated by some authors for meson decays \cite{Maiani:2004uc,Chua:2007cm,Chen:1999nxa}, but since it is completely induced by the non-perturbative QCD effect, the estimate in terms of the present theories cannot be accurate, or at the best can be valid to the order of magnitude if one can find an appropriate model to carry out numerical computations.

\section{Summary}
\label{conclusion}

To evaluate the hadronic matrix elements which are governed by the non-perturbative QCD, phenomenological models
are needed. For the OZI-allowed strong decays,
the QPC model, flux tube model, QCD sum rules and
lattice QCD,  etc. have been successfully used to estimate the decay rates, even though
except the lattice calculation none of them can be directly derived from quantum field theory. We are
assured that all of those models have certain reasonability and they are in parallel somehow.
In this work, we employed the QPC
model to study the strong decay of $Y(4630)\to\Lambda_c\bar\Lambda_c$.

First we assume that $Y(4630)$ is a tetraquark which is a bound state of a diquark and an anti-diquark.
As its mass is slightly above the threshold of two charmed baryons, it would favorably decay into  $\Lambda_c\bar\Lambda_c$ pair, therefore
the fact that $Y(4630)$ is only observed at the invariant spectrum of $\Lambda_c\bar\Lambda_c$, is understandable.

There could be different quantum structures for the diquark-anti-diquark bound state, and we try to assign it with various radial quantum numbers and spin assignments
and then calculate the decay width of
$Y(4630)\to\Lambda_c\bar{\Lambda}_c$ in all possible cases.

The numerical
results show that, within certain parameter range of $R_Y$, one can gain proper decay width $\Gamma_Y$ that agrees with the experimental data if we assign $Y(4630)$ as either
the radial ground state $n_r=1$ or the first radially excited state $n_r=2$.
Whereas for the case of $Y_3$, the obtained partial width are suppressed by the small overlap between the spin wave functions, so the $Y_3$ spin state is ruled out.
Our analysis provides a strong support to the postulate that $Y(4630)$ is the diquark-antidiquark bound state whose mainly decay channel should be $Y(4630)\to\Lambda_c\bar\Lambda_c$.

We are looking forward to getting more information from the Belle-II,
LHCb experiments, especially we will pay more attention to, such as $D\bar{D}$, $D\bar{D}^*$, $D^*\bar{D}^*$, $\psi(2S)\pi^+\pi^-$, $\psi(2S)\eta$ etc, decay modes,
which may shed more light on the structure of $Y(4630)$. In particular,
we suspect if there is a mixing between the tetraquark  and molecular states to result in $Y(4630)$ and $Y(4660)$, it would
be an interesting picture. Indeed
in the near future, with the accumulated data at various
accelerators, our understanding on the $XYZ$ states will be improved
and the observations of  new states are expected.

\bigskip
\noindent{\it Note added.} \ When we make changes to our manuscript, we notice that another work \cite{Sonnenschein:2016ibx} which suggests to use $Y(4630)$ as a window to the landscape of tetraquarks  appears, by J. Sonnenshein and D. Weissman, and we cite it at the end of this modified manuscript.

\section*{Acknowledgement}

We would like to thank Prof. Hai-Yang Cheng for helpful and inspired discussions. We would also
thank  Kan Chen, Yuan Sun and
Hao-Kai Sun who help us with programming for the numerical computations. This project is
supported by the National Natural Science Foundation of China under
Grants No. 11375128 No. 1135009, No. 11222547, No. 11175073. Xiang Liu is also supported by the National
Youth Top-notch Talent Support Program (``Thousands-of-Talents Scheme").

\appendix

\section{Explicit formulae for the matrix elements}
\label{appb}
\begin{equation}
\mathcal{M}^{M_{J_{\Lambda_c}}M_{J_{\bar{\Lambda}_c}}}=-\frac{1}{\sqrt{6}}\gamma\sqrt{8 E_Y E_{\Lambda_c}
E_{\bar{\Lambda}_c}}\mathcal{A}^{M_{J_{\Lambda_c}}M_{J_{\bar{\Lambda}_c}}}
\end{equation}

For spin state $Y_1$:

\begin{eqnarray}
\mathcal{A}^{\frac{1}{2}\frac{1}{2}}&=&\frac{1}{\sqrt{3}} (I^{-1-1}+I^{0-1}+I^{1-1}) \nonumber\\
\mathcal{A}^{\frac{1}{2}-\frac{1}{2}}&=&\mathcal{A}^{-\frac{1}{2}\frac{1}{2}}=-\frac{1}{\sqrt{6}} (I^{-10}+I^{00}+I^{10}) \nonumber\\
\mathcal{A}^{-\frac{1}{2}-\frac{1}{2}}&=&\frac{1}{\sqrt{3}} (I^{-11}+I^{01}+I^{11})
\end{eqnarray}

For spin state $Y_2$:

\begin{eqnarray}
\mathcal{A}^{\frac{1}{2}\frac{1}{2}}&=&-\frac{1}{3} (I^{-1-1}+I^{-10}+I^{00}+I^{1-1}) \nonumber\\
\mathcal{A}^{\frac{1}{2}-\frac{1}{2}}&=&\mathcal{A}^{-\frac{1}{2}\frac{1}{2}}=\frac{1}{3\sqrt{2}}(I^{01}+2I^{1-1}+I^{0-1}) \nonumber\\
\mathcal{A}^{-\frac{1}{2}-\frac{1}{2}}&=&\frac{1}{3} (I^{-11}-I^{00}-I^{10}-I^{11})
\end{eqnarray}

For spin state $Y_3$:

\begin{eqnarray}
\mathcal{A}^{\frac{1}{2}\frac{1}{2}}&=&-\frac{1}{9} (I^{-1-1}+I^{0-1}+I^{1-1}) \nonumber\\
\mathcal{A}^{\frac{1}{2}-\frac{1}{2}}&=&\mathcal{A}^{-\frac{1}{2}\frac{1}{2}}=\frac{1}{9\sqrt{2}} (I^{-10}+I^{00}+I^{10}) \nonumber\\
\mathcal{A}^{-\frac{1}{2}-\frac{1}{2}}&=&-\frac{1}{9} (I^{-11}+I^{01}+I^{11})
\end{eqnarray}

For spin state $Y_4$:

\begin{eqnarray}
\mathcal{A}^{\frac{1}{2}\frac{1}{2}}&=&\frac{1}{9\sqrt{5}} (I^{-1-1}+3I^{-10}-2I^{0-1}-3I^{00}+7I^{1-1}) \nonumber\\
\mathcal{A}^{\frac{1}{2}-\frac{1}{2}}&=&\mathcal{A}^{-\frac{1}{2}\frac{1}{2}}=\frac{1}{9\sqrt{10}} (2I^{-10}-3I^{0-1}-4I^{00} \nonumber\\
&-& 3I^{01}+6I^{1-1}+2I^{10}) \nonumber\\
\mathcal{A}^{-\frac{1}{2}-\frac{1}{2}}&=&\frac{1}{9\sqrt{5}}(I^{11}+3I^{10}-2I^{01}-3I^{00}+7I^{1-1})
\end{eqnarray}

\section{Wave functions}
\label{appa}
In this work, we employ the SHO wave functions for $Y(4630)$ as the input wave functions. For the decay
channels of interest, we need a P-wave two-body wave function for the $Y(4630)$.

For the two-body wave function with quantum numbers $n_r$ and $l$ \cite{Sun:2009tg}
\begin{eqnarray}
\Psi_{n_r=1,l=1}(\mathbf{k})&=&-i2\sqrt{\frac{2}{3}}\frac{R^{5/2}}{\pi^{1/4}}\mathcal{Y}_{1m}(\mathbf{k})\exp\left(-\frac{R^2\mathbf{k}^2}{2}\right),\\
\Psi_{n_r=2,l=1}(\mathbf{k})&=&i\frac{2}{\sqrt{15}}\frac{R^{5/2}}{\pi^{1/4}}(5-2\mathbf{k}^2 R^2)
\mathcal{Y}_{1m}(\mathbf{k})\exp\left(-\frac{R^2\mathbf{k}^2}{2}\right),\nonumber\\
\end{eqnarray}
where   $\mathcal{Y}_{1m}(\mathbf{k})=\sqrt{3/(4\pi)} {\mbox{\boldmath $\epsilon$}}_{-m}\cdot\mathbf{k}$ is the solid harmonic polynomial, with $\epsilon_{\pm1}=(\pm 1/\sqrt{2}, -i/\sqrt{2},0)$ and $\epsilon_{0}=(0, 0,1)$.

\end{document}